\documentclass[seceqn,10pt]{elsart}   

\usepackage{amsmath,amssymb,latexsym,amssymb,tables,psfig,mathrsfs}
\usepackage{macros,subfigure,psfrag,color,bbm}
\usepackage[dvips]{epsfig}
\usepackage[dvips]{graphicx}

\input eqalign.sty

\def\@journal{Physics Letter}
\def\@date{December 2008}

\begin{document}

\begin{frontmatter}

\begin{flushright}
LTH 814 \\
\end{flushright}

\title{Observations on the Wilson fermions in the $\epsilon$ regime}

\author[liv]{Andrea Shindler}
\address[liv]{
Theoretical Physics Division, Dept. of Mathematical Sciences, \\
University of Liverpool, Liverpool L69 3BX, UK
}
\vskip 1.0cm
  \begin{center}
    \includegraphics[draft=false,scale=1]{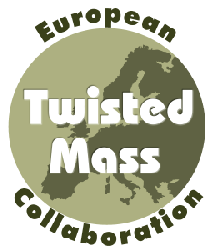}
  \end{center}

\maketitle
\begin{abstract}
We make several observations concerning the low quark mass region
with Wilson fermions and how this is connected with
the $\epsilon$ regime in the continuum.
A transition from tiny cutoff effects to rather large
discretization errors would take place in general 
with Wilson fermions if we lower the quark mass at finite
lattice spacing. We argue that these two regions
exhibit rather different behaviours concerning the
coupling between cutoff effects and zero-modes.
We interpolate between these two regimes
adding to the continuum $\epsilon$ regime formul\ae~,
in the spirit of the Symanzik expansion,
the relevant operators parametrising the leading cutoff effects.
We compute the partition function, the chiral condensate, 
scalar and pseudo-scalar correlation
functions. 
The final formulae can be used to fit lattice data 
to extract physical low energy constants, 
and to estimate systematic uncertainties coming from discretization
errors.
Moreover they suggest ways on how to remove these cutoff effects,
the core of which are captured by the continuum zero modes. 
\end{abstract}

\\
\endkeyword

\end{frontmatter}
\cleardoublepage


\section{Introduction}
\label{sec:intro}

Simulations of lattice QCD in the so-called $\epsilon$ regime~\cite{Gasser:1987ah} of the 
chiral expansion allow in principle the extraction of physical parameters, 
like decay constants and electroweak effective couplings.
Up to now almost all the published results of dynamical simulations in the $\epsilon$ regime 
have been always obtained using lattice chiral invariant formulations, 
like overlap~\cite{Hashimoto:2008fc} (and references therein),
and perfect action~\cite{Hasenfratz:2007yj} fermions.
There are many reasons why chiral invariant lattice fermions are
preferable.
Here we mention the possibility of lowering the quark mass at finite lattice
spacing, without encountering stability or metastabilities problems, small cutoff effects
for a wide range of quark masses, O($a^2$) scaling violations
and continuum-like renormalization patterns. 

Contrary to what is widely believed simulations in the $\epsilon$ regime are in principle
not exclusive to lattice actions with exact lattice chiral symmetry.
It is thus interesting to understand how the continuum $\epsilon$ regime
is probed by Wilson fermions, by mean of analytical and numerical tools.
It has been stressed in the past~\cite{Leutwyler:1992yt,Damgaard:2001js}
that topology plays an important role in this extreme regime.
One could still think of a strategy where simulations performed with Wilson-like fermions 
would sample all the topological sectors. This could be an alternative way to the $\epsilon$ regime.
Preliminary and encouraging results of dynamical simulations in the $\epsilon$ regime
using Wilson twisted mass (Wtm) fermions have been presented at the last lattice 
conferences~\cite{Jansen:2007rx,Jansen:2008ru}. Recently it has been shown~\cite{Hasenfratz:2008fg} 
that a suitable algorithm can sample the configuration space of the $\epsilon$ regime 
also with standard Wilson fermions.
In order to properly interpret the results of these simulations, it is important
to understand how Wilson-like fermions probe the low quark mass region at finite volume.
It is important to understand analytically the quark mass, lattice spacing
and volume dependence with Wilson fermions.
In this Letter we perform a step towards this understanding.
Our analysis will concern Wilson fermions, but it can be extended to Wilson twisted mass.

It is well known~\cite{Gasser:1987ah} that in continuum QCD if one sends 
the quark mass to zero keeping the size of the volume fixed,
chiral symmetry is restored, and the dependence on the quark mass of the chiral condensate is smooth.
It is also well known that at finite lattice spacing and infinite volume
Wilson fermions exhibit a peculiar chiral phase diagram~\cite{Aoki:1984qi,Sharpe:1998xm} which has
important consequences for the quark mass dependence of the condensates
and of the would-be continuum Goldstone bosons.

It is interesting and important to understand which mechanism 
takes place at finite lattice spacing, if we lower the quark mass keeping the 
size of the volume fixed.
One way to study this is to include the effects of the non-vanishing 
lattice spacing in the analysis of~\cite{Gasser:1987ah}, 
to address these issues from the chiral effective theory 
point of view.

The Letter is organized as follows: in sect.~\ref{sec:power} we discuss 
the different patterns of symmetry breaking in the continuum and at finite
lattice spacing and the connection with the choice
of the power counting. This will bring us to find a phenomenological
prescription on how to include cutoff effects in our computations.
Using this prescription in sect.~\ref{sec:part_cond} we analyze and compute partition function and chiral condensate,
and in sect.~\ref{sec:twopoint} the scalar and pseudoscalar correlation
functions. A discussion of the results and conclusions will be given in sect.~\ref{sec:remarks}.

\section{Power counting}
\label{sec:power}

The $\epsilon$ expansion in chiral perturbation theory was introduced in 
~\cite{Gasser:1987ah} and further developed in~\cite{Hasenfratz:1989pk,Hansen:1990un,Hansen:1990yg}.
It is interesting to notice that in these papers no mention is made on the topological sectors
of the configuration space of the theory. In this Letter we will only consider the $N = 2$ flavours case.

The basic feature of the $\epsilon$ expansion is that while the zero modes
are treated non-perturbatively, the non-zero modes fluctuate in a Gaussian way around them and are
treated in a standard perturbative way. This is needed to describe correctly the collective
behaviour of the zero modes when the quark mass is sent to zero at fixed finite volume.
To obtain this in an algebraic way the standard $p$ power counting is modified
treating the would be Goldstone boson mass small compared with the box size.
The standard power counting in the $\epsilon$ regime in the continuum reads
\be
\frac{1}{T} = {\rm O}(\epsilon) , \quad \frac{1}{L} = {\rm O}(\epsilon) , \quad m = {\rm O}(\epsilon^4),
\ee
where $m$ indicates the quark mass and $L$ and $T$ the spatial and temporal extent
of the volume $V=L^3 \times T$.

It is pedagogical to understand which are the lattice parameters which 
ought to be used to reach this regime.
Let us take as a typical range of volumes for dynamical simulations with Wilson fermions
$1.5 {\rm fm} \lesssim L \lesssim 3 {\rm fm}$.
The $\epsilon$ regime is reached if $m\Sigma V \lesssim 1$ . 
If we take the reasonable value of $\Sigma = (250 {\rm MeV})^3$
one easily obtains that, if $T=L$, the value of the properly renormalized 
quark mass should be of the order of
\bea
m &\simeq& 15\quad{\rm Mev} \qquad {\rm for} \qquad  L=1.5\quad{\rm fm} \\
m &\simeq& 6\quad{\rm Mev} \qquad {\rm for} \qquad  L=2\quad{\rm fm} \\
m &\simeq& 2\quad{\rm Mev} \qquad {\rm for} \qquad  L=3\quad{\rm fm}. 
\eea
While this is just an order of magnitude estimate, it is clear that 
quark masses significantly lower than $20$ MeV are needed to be in the $\epsilon$
regime with the current available volumes.

It is interesting now to understand, given these values, and given the current available
lattice spacings which is correct power counting to adopt in describing 
the lattice data.
In fact one can include the effect of the discretization errors in the chiral effective
theory~\cite{Sharpe:1998xm} using as a starting point the continuum 
Symanzik action, describing the interactions of quarks and gluons
with momenta much smaller than $\pi/a$. 
The symmetry properties of the Symanzik action allow to write a generalized chiral expansion
with explicit factors of the lattice spacing $a$.
This has been done in the past in the $p$ regime with great success
for a set of lattice actions~\cite{Bar:2003mh,Aubin:2003mg,Munster:2004am,Sharpe:2004ny} 
and for different power countings.
The most widely used power countings are defined by the way the
quark mass and the lattice spacing are related to each other.
The so-called GSM (generically small masses) regime~\cite{Sharpe:2004ny} is defined by
$m \sim a\Lambda^2$ while the Aoki regime~\cite{Bar:2003mh}, also called large cutoff effects
regime (LCE)~\cite{Aoki:2008gy}, is defined by $m \sim a^2\Lambda^3$.
For illustrative purposes let us take the reasonable value $\Lambda = 250 {\rm MeV}$.
Nowadays dynamical lattice simulations with Wilson fermions are performed
in the range $0.08 {\rm fm} \lesssim a \lesssim 0.04 {\rm fm}$.
This gives us the following values
\bea
a\Lambda^2 &\simeq& 25 \quad {\rm MeV} \qquad a^2\Lambda^3 \simeq 3  \quad {\rm MeV} \qquad  {\rm for }\qquad a=0.08\quad {\rm fm} \\ 
a\Lambda^2 &\simeq& 12 \quad {\rm MeV} \qquad a^2\Lambda^3 \simeq 1  \quad {\rm MeV} \qquad  {\rm for }\qquad a=0.04\quad {\rm fm}.
\eea
A good simulation setup would be with $a=0.04$ fm, a lattice box
of $L/a=T/a=48$ and a quark mass $m=6$ MeV. In this rather ideal case
we would be in a region of quark masses in the GSM regime or quite close to it.
Certainly not everyone will have access to such high quality gauge configurations, 
and moreover it could be advisable anyhow to have more lattice spacings 
to check for scaling violations.
If we would have to increase the value of the lattice spacing,
we would move towards the LCE region possibly getting dangerously close to it,
In fact in the less ambitious setup where $a=0.08$ fm, the lattice box
is $L/a=T/a=24$ and the quark mass $m=6$ MeV we would be quite close to the LCE region.
Even if this discussion is quite simple it gives the hint that in the near future 
if we want to probe the $\epsilon$ regime with Wilson fermions, we
will simulate in the GSM regime but going closer to the LCE regime 

It is thus very important to understand the behaviour of Wilson fermions for small quark masses
in a finite volume.

\subsection{Generic small masses}
\label{ssec:gsm}

In infinite volume the GSM power counting is such that 
$m$ and $a$ are of the same order, as $p^2$.
With this power counting the leading order (LO) chiral Lagrangian is
\be
\mathcal{L}_{W\chi}^{(2)} = \frac{F^2}{4}\Big\{ 
\Tr\left[ \partial_\mu U(x)^\dagger \partial_\mu U(x) \right] - 2B_0
\Tr\left[ \mcM^\dagger U(x) + \mcM U(x)^\dagger \right] - 2aW_0
\Tr\left[U(x) + U(x)^\dagger\right]\Big\},
\label{eq:Wchi}
\ee
where $F$ and $B_0$ are the LECs appearing at LO,
and $W_0$ is an unknown dimensionful constant which parametrises the leading cutoff effects,
$a$ being the lattice spacing.
$\mcM$ is the quark mass matrix and U is the field collecting the Goldstone bosons fields.
We recall that if the $c_{\small{\rm SW}}$ coefficient would be set to its ``correct'' 
non-perturbative value we would have $W_0 = 0$. This does not mean that all the O($a$) terms will disappear, 
because there will be O($a$) terms at higher orders in the chiral expansion, that would have to be, if needed, cancelled by other 
improvement coefficients.
The leading O($a$) can be reabsorbed in the definition of the quark mass~\cite{Sharpe:1998xm}
\be
\mcM \rightarrow \mcM' = \mcM + \frac{W_0}{B_0}a.
\ee
Thus with this power counting the leading order (LO) Lagrangian is identical to the continuum  
LO Lagrangian.

Let us briefly recall why one integrates exactly over the zero modes in the $\epsilon$ 
regime in the continuum.
If one considers the tree-level pion propagator
\be
G(x) = \frac{1}{V}\sum_p \frac{{\rm e}^{ipx}}{p^2 + M_\pi^2} = \frac{1}{V M_\pi^2} + 
\frac{1}{V}\sum_{p\neq 0} \frac{{\rm e}^{ipx}}{p^2 + M_\pi^2} ,
\ee
one sees immediately that the zero mode contribution explodes in the chiral limit
at fixed finite volume.
The standard $p$ expansion would fail in this situation 
because it treats all the modes in the same way. Alternatively one can say that in the
massless limit the zero modes contribution does not appear in the quadratic approximation of the action. 
To circumvent this problem one should reorder the expansion in order to have a representation
of the chiral effective theory with an arbitrary number
of zero modes propagators. This is achieved treating the constant zero modes
in a collective non-perturbative way, and the non-zero modes as standard perturbations~\cite{Gasser:1987ah}.

This tells us that at finite lattice spacing in the GSM regime the LO behaviour of the zero modes 
is like in the continuum, and to correctly include the zero modes propagation in the computation one would
have to treat them in the same way, i.e.
the power counting for GSM in the $\epsilon$ regime is
\be
m = O(\epsilon^4), \quad \frac{1}{L} = O(\epsilon), \quad
\frac{1}{T} = O(\epsilon) \quad a = O(\epsilon^4).
\ee
To understand the possible modifications induced by the discretization errors at NLO
we have to scrutinize the NLO chiral Lagrangian of the GSM regime 
given for example in refs.~\cite{Bar:2003mh,Sharpe:2004ny}.
The NLO Lagrangian has a continuum part $\mcL_{\chi}^{(4)} $ and a part
coming from the discretization effects which reads
\bea 
\mcL_{W\chi}^{(4)} &=& \mcL_{\chi}^{(4)} + a\widetilde{W} \Tr(\partial_\mu U^\dagger \partial_\mu U)\Tr(U + U^{\dagger})
- 2aB_0W \Tr(\mcM'^{\dagger} U + U^\dagger\mcM') \Tr( U + U^{\dagger}) + \nonumber \\
&-& a^2W' \big[\Tr( U + U^{\dagger})\big]^2 - 2aB_0H' \Tr( \mcM'+ \mcM'^\dagger ).
\eea
The notation used here is the same used in ref.~\cite{Sharpe:2004ny}, but here we redefine the LECs 
absorbing the $W_0$ term. As a result their dimension is different and 
given by $\left[W,\widetilde{W}\right] = \left[{\rm Energy}^3 \right]$, 
and $\left[W'\right] = \left[{\rm Energy}^6 \right]$. 
We also do not include all the terms needed for
the analysis of the axial and vector currents which 
we do not consider in this Letter.
The terms of the kind O($am$) ($W$) and O($a^2$) ($W'$) are of order O($\epsilon^4$)\footnote{In the $\epsilon$ expansion
one has to consider the action including the space-time integration which gives a volume 
factor of O($\epsilon^{-4}$).}.
The same is true for the terms of O($ap^2$) ($\widetilde{W}$). 

To explain this let us consider the standard parametrisation for the Goldstone field 
$U$ in the $\epsilon$ expansion
\be
U(x) = U_0\exp\left[\frac{i\pi(x)}{F}\right]
\label{eq:goldstone}
\ee
where $\pi = \pi^a \tau^a$ and the $\tau$ are normalized such that 
$\left\{\tau^a,\tau^b\right\} = 2 \delta^{ab}$.

In the $\epsilon$ expansion the fluctuations around the constant zero modes are treated as O($\epsilon$), 
thus each power of the derivative (or of the momentum $p$) carries at least one power of the fluctuation.
As a result the O($ap^2$) terms are also of O($\epsilon^4$). 

The conclusion is that in the GSM regime the $\epsilon$ expansion with Wilson fermions
is like in the continuum up to O($\epsilon^4$), i.e. if we are in a region of parameters
where we can neglect NNLO terms, Wilson fermions are automatically O($a$) improved.
This is not so surprising for the following reason. In the $\epsilon$ expansion in the continuum the O($\epsilon^0$) and O($\epsilon^2$)
contributions can be computed considering only the LO Lagrangian. In this particular
region contributions from what are called NLO LECs' in the $p$ regime, are suppressed by one order compared to
the standard $p$ regime expansion. 

\subsection{Large cutoff effects region}
\label{ssec:lce}

If we now consider coarser lattice spacings we enter in the
LCE regime and the power counting changes such that
$m'$ and $a^2$ terms are of the same order, as $p^2$.
Already at leading order the situation is different.
The LO Lagrangian is
\be
\mathcal{L}_{W\chi}^{(2)} = \frac{F^2}{4} 
\Tr\left[ \partial_\mu U(x)^\dagger \partial_\mu U(x) \right] - \frac{\Sigma}{2}
\Tr\left[ \mcM'^\dagger U(x) + \mcM' U(x)^\dagger \right] - a^2W'
\left[ \Tr \left( U(x) + U(x)^\dagger\right)\right]^2.
\label{eq:Wchi2}
\ee
where $W'$ is a LEC which parametrises O($a^2$) cutoff effects. It particular its
value depends on the value of $c_{\rm SW}$ adopted but it does not vanish
if the theory is non-perturbatively improved.
In infinite volume the Lagrangian~\eqref{eq:Wchi2} implies a competition of the mass term and of the
O($a^2$) term in the shape of the potential that causes a non-trivial vacuum
structure.
Minimizing the potential gives rise to two possible 
scenarios~\cite{Sharpe:1998xm,Munster:2004am,Scorzato:2004da,Sharpe:2004ps} 
for the phase diagram of Wilson-like fermions: \\
\vspace{-0.6cm}
\begin{itemize}
\item the Aoki scenario $W'<0$~\cite{Aoki:1984qi}; \\
\item the Sharpe-Singleton scenario $W'>0$~\cite{Sharpe:1998xm}.
\end{itemize}
Depending on the sign of $W'$ the two scenarios predict a different 
pattern for the quark mass dependence of the chiral condensate and of the 
pion masses. In the Sharpe-Singleton scenario the pion masses do not go to zero
in the chiral limit, while in the Aoki scenario within the Aoki phase the charged
pion stay massless while the neutral pion mass becomes massive.
Recalling our discussion in the previous section this means that for the 
Sharpe-Singleton scenario the pion mass will never vanish, and as a consequence one will still have
the zero-modes in the quadratic approximation. 
This implies that in this scenario a standard perturbative expansion
can be performed and no collective phenomena take place~\cite{Sharpe:2008ke}.
In the Aoki scenario, the Goldstone boson manifold is not $SU(2)$ like in the continuum, 
and an appropriate expansion should be performed taking into account 
the pattern of symmetry breaking at finite lattice spacing~\cite{Sharpe:2008ke}.
This does not mean that the $\epsilon$ regime does not exist for Wilson fermions.
In fact for example if we stay in the GSM regime, like we have discussed in the previous section,
one could simulate in the $\epsilon$ regime being affected by very tiny cutoff effects.
It simply means that the order of the chiral and continuum limit has to be understood with care.

From this discussion it is clear that 
Wilson fermions in the low quark mass region regime have a transition
in terms of cutoff effects from no cutoff effects (up to NNLO) in the GSM regime
to a regime where cutoff effects appear at LO and the collective zero-modes
phenomena, if present, are pretty different from the continuum.

To understand up to which value of the lattice spacing and quark mass
we can still discuss of continuum zero modes we can
imagine a situation where we have continuum-like collective phenomena
and the Lagrangian of the LCE regime~\eqref{eq:Wchi2}. While we know this is not
a completely consistent procedure it might give us a hint 
to where this picture breaks down.

The power counting would be
\be
m' = O(\epsilon^4), \quad \frac{1}{L} = O(\epsilon), \quad
\frac{1}{T} = O(\epsilon) \quad a^2 = O(\epsilon^4).
\ee
and we expect to encounter some contradictions, indicating that we cannot arbitrarily
lower the quark mass at fixed lattice spacing, at least if we insist in
integrating over continuum-like zero modes.

Expanding the LO Lagrangian we obtain for the action the following decomposition
\be
\Slo = \act{\Slo}{A} + \act{\Slo}{B} + \act{\Slo}{C},
\label{eq:Swchi}
\ee
where 
\be
\act{\Slo}{A} = -\frac{\Sigma}{2}\int d^4x \Tr \left[\mcM'^{\dagger} U_0+ U_0^{\dagger} \mcM'\right], \qquad \act{\Slo}{B} = \frac{1}{4}\int d^4x \Tr \left[\partial_\mu \pi(x) \partial_\mu \pi(x) \right]
\label{eq:Scont}
\ee
\be
\act{\Slo}{C} = -W'a^2\int d^4x   \left[ \Tr \left(U_0 + U_0^{\dagger} \right) \right]^2.
\ee
If we now use standard techniques we can write the partition function up to
a normalization factor as
\be
\mcZ = \mcA \int_0^\pi d \theta \sin^2\theta\e^{z_1\cos\theta + z_2\cos^2\theta} = 
\mcA \int_0^\pi d \theta \sin^2\theta\e^{z_1\cos\theta}
\sum_{n=0}^\infty \frac{z_2^n}{n!}(1-\sin^2\theta)^n
\ee
where assuming a mass matrix proportional to the identity we define
\be
z_1 = 2m'\Sigma V, \qquad z_2 = 16 a^2V W'.
\ee
We see here the towers of infinite O($a^2$) terms induced by the coupling of the 
zero modes with the cutoff effects.
Using the binomial formula we can write 
\be
\mcZ = \mcA \sum_{n=0}^\infty \frac{z_2^n}{n!} \sum_{k=0}^n\binom {n} {k} (-1)^k 
X_{k+1}(z_1)
\ee
where we have defined the function
\be
X_{k+1}(z_1) = \frac{\sqrt{\pi}\Gamma(k+3/2)}{(z_1/2)^{k+1}}I_{k+1}(z_1)
\label{eq:Xk}
\ee
and $I_{k+1}(z_1)$ are modified Bessel functions.
To understand where this kind of expansion breaks down 
we perform the infinite volume limit at fixed lattice spacing and 
fixed quark mass trying to reproduce 
the same pattern of the results obtained in infinite volume~\cite{Sharpe:1998xm}. 
So we want to study the behaviour of the partition function for large $z_1$
and $z_2$ keeping their ratio fixed. The asymptotic expansion 
of the Bessel functions at leading order in $1/z_1$ reads
\be
I_{k+1}(z_1) \sim\frac{\e^{z_1}}{\sqrt{2\pi z_1}},
\ee
so for large $z_1$ we have
\be
\mcZ \sim \mcA \sum_{n=0}^\infty \frac{z_2^n}{n!} \sum_{k=0}^n\binom {n} {k} (-1)^k 
\frac{\sqrt{\pi}\Gamma(k+3/2)}{(z_1/2)^{k+1}}\frac{\e^{z_1}}{\sqrt{2\pi z_1}}.
\ee
The series can be rearranged in the following form
\be
\mcZ \sim \mcA \frac{2\e^{z_1+z_2}}{\sqrt{2 z_1^3}} \sum_{k=0}^\infty \left(\frac{-2z_2}{z_1}\right)^k
\frac{\Gamma(k+3/2)}{k!} 
\ee
The radius of convergence of this power series with argument $\left|\frac{2z_2}{z_1}\right|$ is given by
\be
\lim_{k \rightarrow \infty} \left|\frac{\Gamma(k+3/2)}{\Gamma(k+5/2)}\frac{(k+1)!}{k!}\right|
\ee
which given the properties of the $\Gamma$ function is $1$.
We conclude that if we take as a starting point the action in~\eqref{eq:Swchi} 
at fixed lattice spacing and quark mass for large volumes the partition function 
is well defined only if 
\be
z_1 > 2|z_2|, \qquad {\rm with} \qquad z_1 > 0 
\label{eq:bound}
\ee
We recall that the point $z_1 = 2|z_2|$ is exactly the location
of the phase transition in the Aoki scenario. This implies that
if we want to study the $\epsilon$ regime with W$\chi$PT retaining
the continuum-like integration over the zero modes we cannot
lower the quark mass beyond the border of the phase transition.
Moreover it is conceivable that given the strong coupling of the zero-modes with
the cutoff effects it is better, to avoid large cutoff effects, to stay not so close 
to the phase transition point.

The bound given in eq.~\eqref{eq:bound} is obtained integrating over the 
continuum zero modes, thus it suggests that if we want to go below the bound we 
have to change the way we integrate over the zero modes.

Equivalently in the infinite volume limit we are able to reproduce
the well-known results concerning the chiral phase diagram of Wilson fermions~\cite{Sharpe:1998xm}
for quark masses not lower than the location of the Aoki phase transition.
If we would like to go beyond this limit we would need to
parametrize our integration over the zero modes taking into account 
the nature of the spontaneous symmetry breaking (SSB) taking place in infinite volume~\cite{Sharpe:2008ke}.

If we want to keep the description in terms of continuum-like zero modes
one needs stay in the GSM regime or at most between the GSM and the LCE regime.

\section{Partition function and chiral condensate}
\label{sec:part_cond}

In this section we concentrate on the transition region between the GSM regime
and the LCE regime.
We try to understand the impact of the cutoff effects with Wilson
fermions in the $\epsilon$ regime if we simulate between
the GSM regime and the LCE regime treating the operators parametrising
the cutoff effects as insertions.

From the power counting point of view this can be seen in two ways, 
either as performing the computation
up to O($\epsilon^2$) using the GSM power counting 
and including terms of O($\epsilon^4$) containing 
the relevant operator parametrising the cutoff effects, or performing
the computation up to O($\epsilon^2$) in the LCE, but treating the 
O($a^2$) cutoff effects terms as small compared to the mass term.
In both cases the pattern of spontaneous symmetry breaking is as
in the continuum.
In our analysis we want to keep O($a^2$) and O($am$) terms. 
Keeping the O($a^2$) term is equivalent of retaining the first term
of the expansion done in the previous section which interpolates between the two regimes.
We add also the O($am$) to understand the impact of this term in the correlation functions.

We have already seen in sec.~\ref{ssec:gsm} that 
the LO Lagrangian in the GSM regime is like in the continuum
\be
\mathcal{L}_{W\chi}^{(2)} = \frac{F^2}{4} 
\Tr\left[ \partial_\mu U(x) \partial_\mu U(x)^\dagger \right] - \frac{\Sigma}{2}
\Tr\left[ \mcM'^\dagger U(x) + \mcM'U(x)^\dagger\right],
\label{eq:Wchi2b}
\ee
as the pattern of spontaneous symmetry breaking.
We can thus use the standard parametrisation for Goldstone fields~\eqref{eq:goldstone}.
Expanding the LO Lagrangian we obtain for the action the following decomposition
\be
\Slo = \act{\Slo}{A} + \act{\Slo}{B},
\ee
where $\act{\Slo}{A,B}$ are defined in eq.~\eqref{eq:Scont},
which are the standard LO continuum contribution from the zero modes.

At the next order we have two types of contributions coming from the expansion to O($\epsilon^2$) of the LO Lagrangian 
and from the cutoff effects.
We write then the next term of the action as 
\be
S_4 = \Slon + \Snlo
\ee
where 
\be
\Slon = \act{\Slon}{A} + \act{\Slon}{B}
\ee
is the standard O($\epsilon^2$) contribution in the continuum
\be
\act{\Slon}{A} = \frac{1}{48 F^2}\int d^4x \Tr \left\{ \left[\partial_\mu \pi,\pi\right] \left[\partial_\mu \pi,\pi\right] \right\}
\ee
\be
\act{\Slon}{B} = \frac{\Sigma}{4F^2}\int d^4x \Tr \left[\mcM' U_0\pi^2 + \pi^2U_0^{\dagger} \mcM' \right]
\ee
while
\bea
\Snlo &=& \int d^4x \left\{ -W'a^2 \left[\Tr\left(U_0 + U_0^\dagger\right) \right]^2 +
                \frac{a\widetilde{W}}{F^2}\Tr\left[\partial_\mu \pi \partial_\mu \pi \right]\Tr\left[U_0 + U_0^{\dagger}\right] + \right. \nonumber \\
         &-& \left. \frac{2aW\Sigma}{F^2}\Tr\left[\mcM'^{\dagger}U_0 + U_0^{\dagger}\mcM'\right] \Tr\left[U_0 + U_0^{\dagger}\right] - 
                \frac{2a\Sigma H'}{F^2}\Tr\left[\mcM'^\dagger + \mcM'\right]\right\}.
\eea
These are the terms we want to include containing only lattice artifacts. 
The last term is responsible for the contact term in the chiral condensate.
In the following we will call 
the inclusion of $\Slon$ and $\Snlo$ an O($\epsilon^2$) expansion keeping in mind
that includes few O($\epsilon^4$) terms in the GSM power counting.

The measure of the partition function can be splitted with appropriate Jacobian $J\left(\pi\right)$
\be
\mcZ = \int \du \e^{-S\left[U\right]} = \int\duzero \dpi J\left(\pi\right) \e^{-S\left[U\right]}
\ee
and written in the following exact form
\be
\mcZ = \int \duzero \e^{-\act{\Slo}{A}} \mcZ_\pi \left[U_0\right],
\label{eq:z}
\ee
where
\be
\mcZ_\pi\left[U_0\right] = \int \dpi J\left(\pi\right) \e^{-S\left[U\right] +\act{\Slo}{A}}.
\label{eq:zpi}
\ee
To expand the partition function up to O($\epsilon^2$) we need to expand the Jacobian and the
exponential in eq.~\eqref{eq:zpi} up to O($\epsilon^2$). 
For convenience we rewrite the LO partition function as 
\be
\mcZ^{(0)} = \mcZ_0^{(0)}\mcZ_\pi^{(0)},\qquad {\rm where} \qquad 
\mcZ_0^{(0)} = \int \duzero \e^{-\act{\Slo}{A}}, \qquad 
\mcZ_\pi^{(0)} = \int \dpi \e^{-\act{\Slo}{B}}.
\ee
Expanding to O($\epsilon^2$) $Z_\pi\left[U_0\right]$ we obtain terms which depends on $U_0$
like $\act{\Slon}{B}$ and terms that, like the Jacobian, do not. These latter will enter only in the
absolute normalization of the partition functions. We then end up with several terms 
\bea
Z_\pi\left[U_0\right] &=& \mcN\left\{1 - 
\frac{\Sigma V \left(N_f^2-1\right)}{F^2N_f}\Tr\left[\mcM^{\dagger}U_0 + U_0^{\dagger}\mcM\right]\bar{G}(0) + \right. \nonumber \\ 
&+& \left.W'a^2V\left[\Tr\left(U_0+U_0^\dagger\right)\right]^2 + 
\frac{2 a W\Sigma V}{F^2}\Tr\left[\mcM^\dagger U_0 + U_0^\dagger\mcM\right]\Tr\left[U_0 + U_0^\dagger\right] + \right. \nonumber \\
&+& \left. \frac{2a\Sigma H'}{F^2}\Tr\left[\mcM'^\dagger + \mcM'\right]\right\},
\label{eq:zpinlo}
\eea
where $\bar{G}(x-y)$ is the 'pion' propagator without the contribution
of the zero mode defined by
\be
\frac{1}{Z_\pi^{(0)}}\int \dpi \e^{-\act{\Slo}{B}}\pi_{ij}(x)\pi_{kl}(y) = 2\left(\delta_{il}\delta_{kl} - \frac{1}{N_f}\delta_{ij}\delta_{kl}\right)\bar{G}(x-y).
\ee
In dimensional regularization the propagator $\bar{G}(0)$ is finite and is given by
\be
\bar{G}(0) = -\frac{\beta_1}{\sqrt V}, \qquad {\rm with} \qquad V=L^3\times T
\ee
and $\beta_1$ is a numerical constant which depends only on the geometry of the box~\cite{Hasenfratz:1989pk}.

In eq.~\eqref{eq:zpinlo} the first term that can be reabsorbed in the definitions of $\Sigma$, thus defining 
an effective LEC $\Sigma_{\rm eff} = \Sigma \rho$ 
where 
\be
\rho = 1-\frac{N_f^2-1}{N_fF^2}\bar{G}(0).
\ee
The partition function at O($\epsilon^2$) can now be written as 
\bea
\mcZ &=& \mcN \int \duzero \e^{-\act{\Slo}{A}\left[\Sigma_{\rm eff}\right]} \times 
\left\{1 + W'a^2V \left[\Tr\left(U_0+U_0^\dagger\right)\right]^2 + \right. \nonumber \\
&+& \left. \frac{2 a W\Sigma V}{F^2}\Tr\left[\mcM^\dagger U_0 + U_0^\dagger\mcM\right] \Tr\left[U_0 + U_0^\dagger\right] + 
\frac{2a\Sigma H'}{F^2}\Tr\left[\mcM'^\dagger + \mcM'\right]\right\},
\label{eq:znlo}
\eea

We can now analyze various features of this formula. The first thing to notice is that with the
choice of our power counting the Boltzmann factor is as in the continuum, while cutoff effects
appear as standard perturbative corrections.
In the continuum the O($\epsilon^2$) contribution to the partition function is reabsorbed in the definition of $\Sigma$. 
Here at finite lattice spacing we define have additional terms that cannot be reabsorbed. 

We can now proceed and compute the chiral condensate.
It is well known that the direct computation of the chiral condensate
is a difficult task with Wilson fermions. Here we simply use it as an
example to explain the computational procedure.
We first define the quark currents in a standard fashion
\be
S^0(x) = \psibar(x)\psi(x), \qquad P^0(x) = \psibar(x)i\gamma_5\psi(x)
\ee
\be
S^a(x) = \psibar(x)\frac{\tau^a}{2}\psi(x), \qquad P^a(x) = \psibar(x)i\gamma_5\frac{\tau^a}{2}\psi(x)
\ee
where $\psi$ is a flavour doublet quark field. 

To compute correlation functions in the effective theory we use the
standard method of augmenting the mass term with sources for scalar and pseudoscalar
currents
\be
\mcM \longrightarrow \mcM + s(x) + ip(x)
\ee
where 
\be
s(x) = s^0(x) + s^a(x)\frac{\tau^a}{2}, \qquad p(x) = p^0(x) + p^a(x)\frac{\tau^a}{2}.
\ee
The corresponding correlation functions can now be obtained performing functional derivatives
respect to the partition function~\eqref{eq:z}, which now depends on the sources, and then setting the sources to zero.

We can now write the scalar and pseudoscalar current at NLO in the effective theory
\be
-S^0(x) = \frac{\Sigma}{2}\Tr\left[U(x) + U(x)^\dagger\right] + 
\frac{2a\Sigma W}{F^2}\left[\Tr\left(U(x) + U(x)^\dagger\right)\right]^2 + \frac{4aH'\Sigma N_f}{F^2},
\ee
\be
P^a(x) = \frac{\Sigma}{2}\Tr\left[\frac{i\tau^a}{2}\left(U(x) - U(x)^\dagger\right)\right] + 
\frac{2a\Sigma W}{F^2}\Tr\left[\frac{i\tau^a}{2}\left(U(x) - U(x)^\dagger\right)\right]
\Tr\left(U(x) + U(x)^\dagger\right).
\ee
Expanding this expression to O($\epsilon^2$) and performing the contractions
on the $\pi$ field, in the same way as we did for the partition function, we obtain for the chiral
condensate the following expression
\be
-\langle S^0 \rangle = \frac{\Sigma_{\rm eff}}{2}\langle\Tr\left[U_0 + U_0^\dagger\right]\rangle_2 + 
\frac{2a\Sigma_{\rm eff} W}{F^2}\langle\left[\Tr\left(U_0 + U_0^\dagger\right)\right]^2\rangle_0 + 
\frac{8aH'\Sigma_{\rm eff}}{F^2}.
\ee
where we have defined two different expectation values over the zero modes
\be
\langle \mcO\left[U_0\right]\rangle_0 = \frac{1}{\mcZ_0^{(0)}} \int \duzero  \mcO\left[U_0\right] 
\e^{-\act{\Slo}{A}\left[\Sigma_{\rm eff}\right]}
\ee
and
\be
\langle \mcO\left[U_0\right]\rangle_2 = \langle \mcO\left[U_0\right]\rangle_0 - \langle \mcO\left[U_0\right]\Snlo\left[U_0\right]\rangle_0 +
\langle \mcO\left[U_0\right]\rangle_0\langle \Snlo\left[U_0\right]\rangle_0.
\ee
In the last two terms of this formula we can compute the expectation values not considering the effective LEC 
$\Sigma_{\rm eff}$ but only $\Sigma$, the difference being of higher order. 
We define now the following LECs
\be
\delta_W = \frac{16W}{F^2}, \qquad w' = \frac{16W'}{F^2}.
\ee
This definition is similar to the one done in ref.~\cite{Sharpe:2004ny}, but here we keep explicit the $a$
dependence in the correlation functions and we redefine the LECs absorbing the $W_0$ term.
As a result their dimension is different and given by $\left[\delta_W\right] = \left[{\rm Energy} \right]$
and $\left[w'\right] = \left[{\rm Energy}^4 \right]$.
These LECs have a very precise origin: $\delta_W$ parametrises the mass dependent O($a$) cutoff effects
of the pion mass and $w'$ the O($a^2$) cutoff effects of the pion mass. Setting $c_{\rm SW}$ to its 'correct' 
non-perturbative value sets $\delta_W=0$ but simply changes the value of $w'$\footnote{If we use Wilson twisted mass
fermions the same LECs play a relevant r\^ole: $\delta_W$ parametrises the dependence of the PCAC mass on the twisted mass
close to maximal twist and $w'$ the O($a^2$) pion mass splitting.}.

The final expression for the chiral condensate reads
\bea
\langle S^0 \rangle &=& 2 \Sigma_{\rm eff}\left\{ \frac{X_1'(z)}{X_1(z)} \left[ 1 + z_2 \left(\frac{X_2(z) - X_2'(z)}{X_1(z)}\right)
+ a \delta_W z \left( \frac{X_2(z) - X_2'(z)}{X_1(z)}\right) \right] + \right. \nonumber \\
&+& \left. a \delta_W\left(1 - \frac{X_2(z)}{X_1(z)}\right)
+ \frac{8 a H'}{F^2}\right\}.
\eea
Where $z_2 = a^2w'F^2V$, the argument of the functions is $z=2m\Sigma_{\rm eff} V$ and we have defined
\be
X_1(z) = \frac{\pi}{z}I_1(z), \qquad X_2(z) = \frac{3\pi}{z^2}I_2(z) 
\ee
where $I_1$ and $I_2$ are modified Bessel functions (definitions and properties can be found for example in~\cite{Abramowitz:1964}).
Important checks are obtained performing the infinite volume limit and the continuum limit.
If we send $a \rightarrow 0$ we obtain the O($\epsilon^2$) result of Gasser and Leutwyler~\cite{Gasser:1987ah}.
If we send $V \rightarrow \infty$ we obtain a formula for infinite volume and finite lattice spacing.
This is consistent with the formula given for example in ref.~\cite{Sharpe:2004ny}, if we just compare the terms
we have actually included in our computation. In particular the terms which are proportional to $a^2V$ or
$az$ both vanish in the infinite volume limit because for large $z$
\be
\frac{X_2(z) - X_1(z)}{X_1(z)} \sim \frac{21}{2z^2}
\ee
In the left plot of fig.~\ref{fig:pp} we show the dependence of the chiral condensate on the quark mass, for a lattice spacing given by
$z_2=1$ and a volume of $V=L^3\times 2L$ with $L=2$ fm, where we have subtracted the contact term.
We also use the values for the LECs: $\Sigma = (250 {\rm MeV})^3$, $F=86.2 {\rm Mev}$, $a\delta_W = 0.1$.

\section{Two-point functions}
\label{sec:twopoint}

\begin{figure}
\hskip-0.8cm
\includegraphics[width=0.45\textwidth,angle=270]{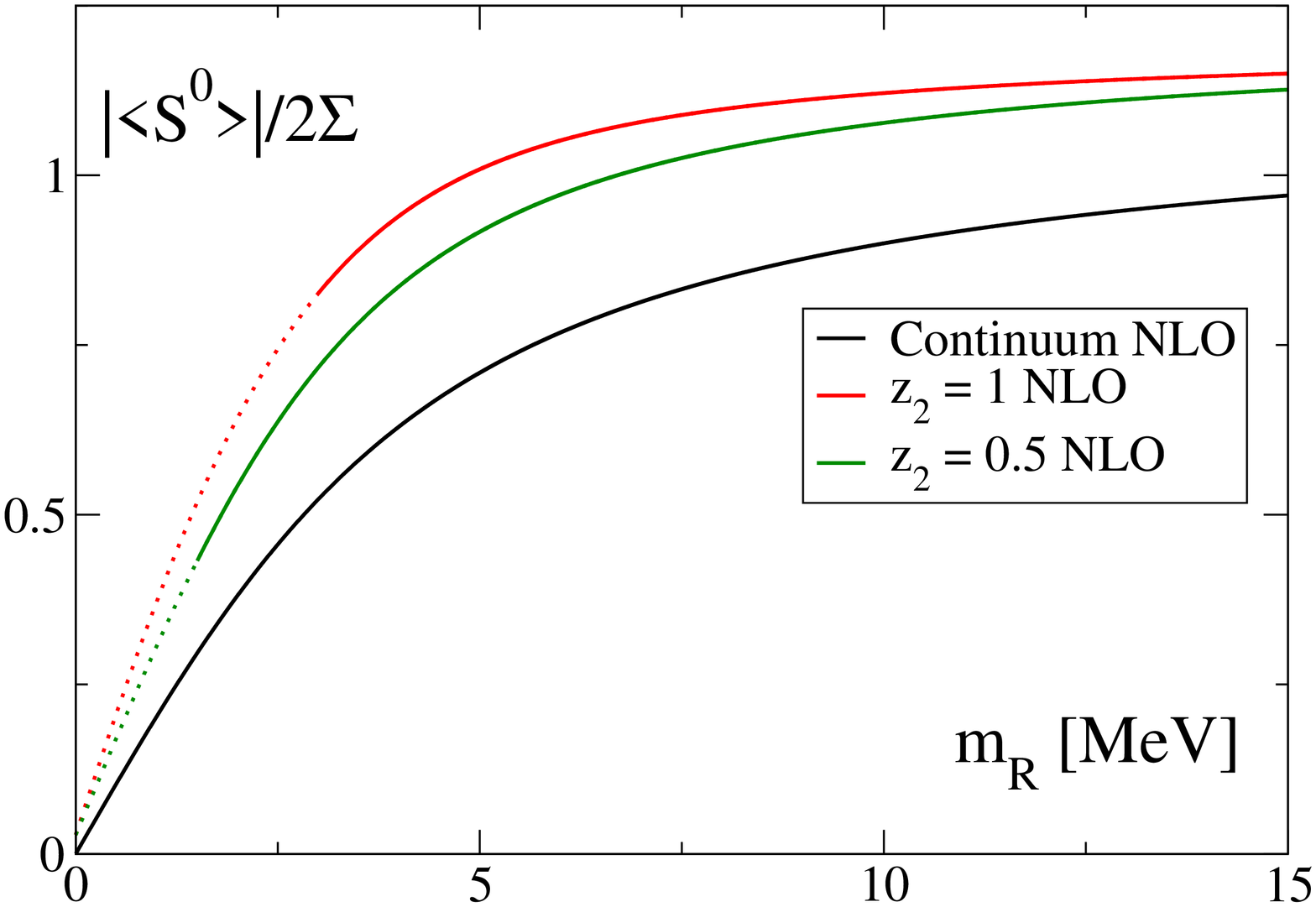}
\includegraphics[width=0.45\textwidth,angle=270]{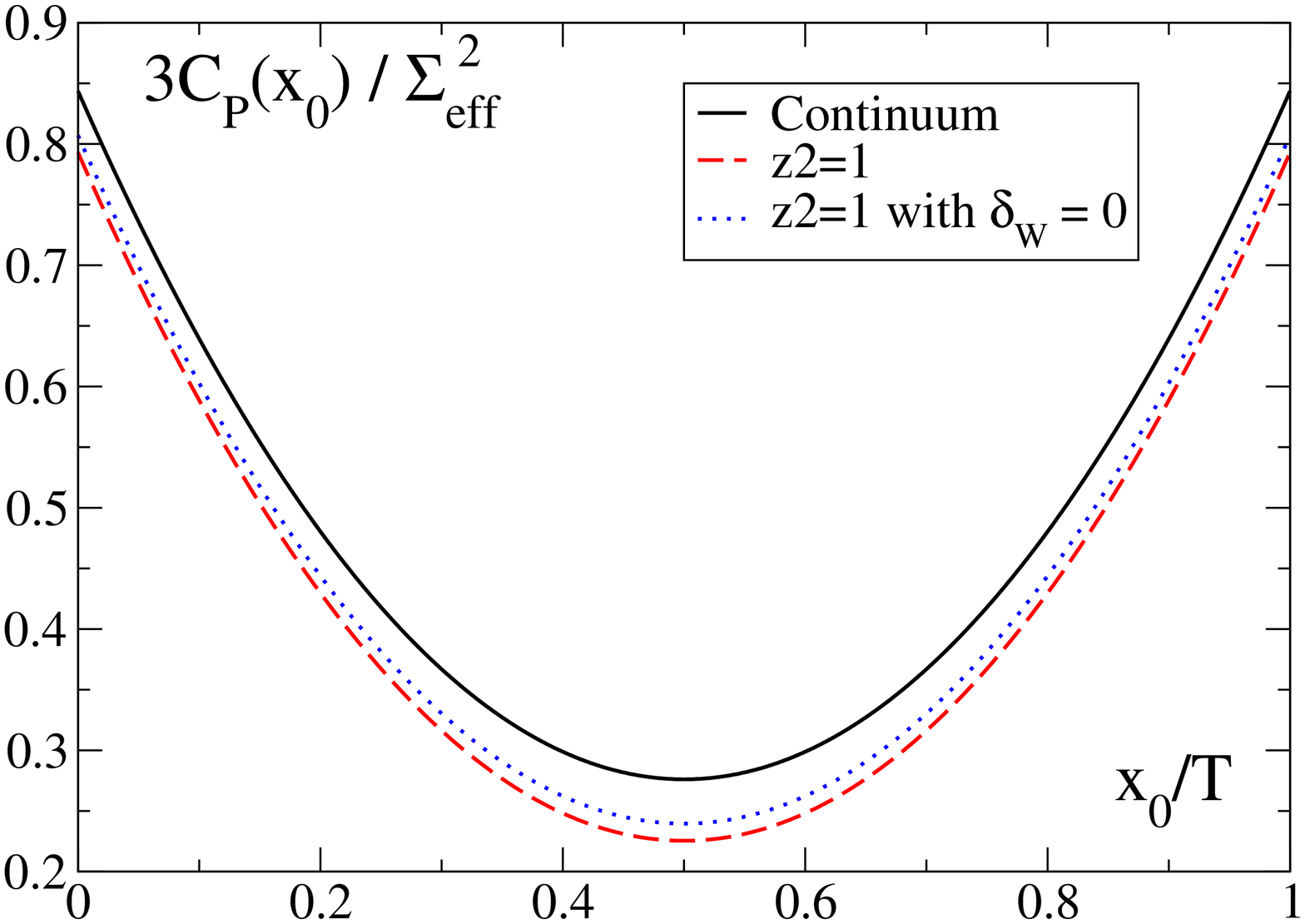}
\caption{Left plot: mass dependence of the subtracted and properly normalized chiral condensate in the continuum and at finite
lattice spacing. The dashed lines indicate the range of quark masses where the expansion carried out in this work
breaks down. This range shrinks to zero towards the continuum limit. 
Right plot: Euclidean time dependence of the pseudoscalar two-point function in the continuum
and at finite lattice spacing with 2 values of $a\delta_W=0,0.1$.}
\label{fig:pp}
\end{figure}

In the previous two sections we have explained in detail the strategy we wish to adopt,
and we can repeat a similar computation for scalar and pseudoscalar two-point functions.
We define 
\be
\frac{1}{L^3}\int d^3 x \langle S^0(x) S^0(0)\rangle = C_{\rm S}(x_0), \qquad 
\frac{1}{L^3}\int d^3 x \langle P^a(x) P^b(0)\rangle = \delta^{ab}C_{\rm P}(x_0),
\ee
and after performing the group integrals we obtain ($X_3(z)$ is defined in eq.~\eqref{eq:Xk})
\bea
C_{\rm S}(x_0) &=& 4\Sigma_{\rm eff}^2\left\{1-\frac{X_2(z)}{X_1(z)} + 
\left( z_2 + a \delta_W z\right) \left[\frac{X_3(z)}{X_1(z)} - \frac{X_2^2(z)}{X_1^2(z)}\right] \right. \nonumber \\
&+& \left. 2a\delta_W\left[ \frac{X_1'(z) - X_2'(z)}{X_1(z)}\right] + \frac{8 a H'}{F^2}\frac{X_1'(z)}{X_1(z)}
+ \frac{1}{F^2}\frac{X_2(z)}{X_1(z)}\frac{T}{L^3}h_1(\tau) \right\}
\label{eq:SS}
\eea
where $\tau=x_0/T$ and where the time dependence of the massless propagator
is given, as in the continuum by~\cite{Hasenfratz:1989pk}
\be
h_1(\tau) \equiv \frac{1}{T}\int d^3 x \bar{G}(x) = \frac{1}{2}\left[ \left(\tau -\frac{1}{2}\right)^2 - \frac{1}{12}\right].
\ee
For the pseudoscalar correlation function we obtain
\bea
C_{\rm P}(x_0) &=& \frac{\Sigma_{\rm eff}^2}{3}\left\{\frac{X_2(z)}{X_1(z)} - 
\left(z_2 + a \delta_W z \right) 
\left[\frac{X_3(z)}{X_1(z)} - \frac{X_2^2(z)}{X_1^2(z)}\right] \right. \nonumber \\
&+& \left. 2a\delta_W\left[ \frac{X_2'(z)}{X_1(z)}\right] + \frac{3}{F^2}\left( 1 - \frac{1}{3}\frac{X_2(z)}{X_1(z)}\right)\frac{T}{L^3}h_1(\tau) \right\}
\label{eq:PP}
\eea
Using the same values for the LECs' we have used for the chiral condensate in the right plot of 
fig.~\ref{fig:pp} we show the Euclidean time dependence of the two-point functions.

Formul{\ae}~\eqref{eq:SS} and \eqref{eq:PP} suggests an interesting way to remove most of the
cutoff effects, which is to study the correlation function
$C_{\rm S}(x_0)/4+3C_{\rm P}(x_0)$.
In fact in this correlation function all the cutoff
effects are absent but the term proportional to $2a\delta_W\left[ \frac{X_1'(z)}{X_1(z)}\right]$.
We can immediately recognize the ratio $\frac{X_1'(z)}{X_1(z)}$ has the 
continuum mass dependence of the chiral condensate in the symmetry restoration
region. So this term is expected to be very small in the $\epsilon$ 
regime and can be removed determining in a non-perturbative way $c_{\rm SW}$.
The correlation function will still have a suitable dependence on the LECs,
so it can still be used to fit lattice data.
The obvious drawback is that $C_{\rm S}(x_0)$ would require the computation
of a disconnected diagram. 

\section{Conclusions and outlooks}
\label{sec:remarks}

In this Letter we have analyzed some aspects of the chiral limit with Wilson fermions.
We have discussed how cutoff effects influence the way we reach the $\epsilon$ regime,
and to which extent we can continue to discuss about zero-modes as 
it is done in the continuum. We have found that concerning cutoff effects 
there is an abrupt transition between the GSM regime, where Wilson fermions
show no cutoff effects up to NNLO, and the LCE regime where cutoff effects
appear at LO directly in the action, in a non-perturbative way. 
This transition is driven by the coupling of the zero modes with
O($a^2$) effects. A first natural way to avoid these cutoff effects is to keep the quark mass heavier
than the values at which one would expect phase transitions in infinite volume,
or better in the GSM regime where Wilson fermions are automatically O($a$) improved
up to NNLO and continuum formul{\ae} could be used. 
One can also simulate in the transition region between the GSM and the LCE regimes. 
For those values of quark masses we can still use W$\chi$PT in the $\epsilon$ regime 
in a very similar way to what is done in the continuum, 
i.e. keeping the integration over the $SU(2)$ manifold for the zero modes,
and treating the cutoff effects as perturbative corrections. 
In this transition region we have computed formul{\ae} which describe the impact of the 
cutoff effects as soon as we move away from the GSM
regime. This formul{\ae} could be used to fit lattice data in order to better
understand the systematics of the simulation results coming from discretization errors.

The analysis presented in this Letter indicates a second way on how to control
the increase of the discretization errors in the transition from the GSM and the LCE regime.
Our results give strong indications that the rise of the cutoff effects 
is driven by the zero modes.
It is possible that, if one is able to remove the zero modes
from the analysis of the lattice data, the cutoff effects will be 
again under control. This could be done for example 
finding a way to define in a robust way
configurations with trivial topology.
The analysis presented here seems to indicate that the cutoff effects blow up
only in sectors with non-vanishing
topological charge, i.e. with zero modes.
\vspace{-1.0cm}
\section*{Note added}
\vspace{-1.0cm}
After this work was completed and sent to the whole ETMC I was informed and then
I received a paper~\cite{Bar:2008th} about Wilson fermions in the $\epsilon$ regime
where similar conclusions have been reached.
\vspace{-1.0cm}
\section*{Acknowledgments}
\vspace{-1.0cm}
I thank Karl Jansen for crucial discussions at the beginning of this work
about the $\epsilon$ regime.
I acknowledge interesting discussions with P. Damgaard, C. Michael and R. Sommer.
I also wish to thank K. Jansen, C. Michael and G. M\"unster for a 
careful reading of the manuscript and all the members of ETMC for a
most enjoyable collaboration.

\bibliographystyle{h-elsevier}    
\bibliography{weps}      
\end{document}